\documentclass[aps,prb,amsmat,amssymb,amsfonts,superscriptaddress,twocolumn,
tightenlines, notitlepage]{revtex4-2}
\usepackage[section]{placeins}
\usepackage{relsize}
\usepackage{amsfonts}
\usepackage{hyperref}
\usepackage{bbm}
\usepackage{mathtools}
\usepackage{tikz}
\usepackage{physics}
\usepackage{amsthm}
\usepackage{amsmath}
\usepackage{amssymb}
\usepackage{tensor}
\usepackage{bigints}
\usepackage{mdframed}
\usepackage{xcolor}
\usepackage{graphicx}
\usepackage{float}
\usepackage{booktabs}
\usepackage{cleveref}
\usepackage{orcidlink}

\newcommand{\qexp}[1]{\langle#1\rangle}

\begin{document}

\title{State Space Geometry of the Spin-1 Antiferromagnetic Heisenberg Chain}
\date{\today}
\author{James Lambert}
\email{lambej3@mcmaster.ca}
\author{Erik S. S{\o}rensen\,\orcidlink{0000-0002-5956-1190}}
\email{sorensen@mcmaster.ca}
\affiliation{Department of Physics \& Astronomy, McMaster University
1280 Main St.\ W., Hamilton ON L8S 4M1, Canada}

\begin{abstract}

We study the phase diagram of the spin-1 antiferromagnetic Heisenberg chain with uniaxial anisotropy and applied magnetic field in terms of the genuine multipartite entanglement as witnessed by the mean quantum Fisher information density. By generalizing  the manifold studied in ~\cite{yu2020experimental,palumbo2018revealing} to the many body case for spin 1, we connect the state space curvature in the vicinity of the ground state of the Heisenberg chain to the genuine multipartite entanglement. Our analysis demonstrates that the quantum critical points and symmetry protected topological (SPT) phase exhibit large state space curvature, while the separable phases are completely flat, offering insight into the physical interpretation of state space curvature. We further show that the entanglement in the SPT phase is enhanced by the presence of uniaxial anisotropy, and undiminished in the presence of uniform magnetic fields. The magnon condensate phase induced by large fields is shown to emanate from the Gaussian critical point, and exhibits massive multipartite entanglement over a robust region of the parameter space. 
    
\end{abstract}

\maketitle
\section{Introduction}

Since the work of Zanardi~\cite{zanardi2006ground} and Gu~\cite{yang2008fidelity}, there has been substantial interest in investigating the properties of many body ground states from the perspective of quantum information geometry (QIG) both theoretically~\cite{zanardi2006ground,zanardi2007bures,zanardi2007information,Albuquerque2010,gabbrielli2018multipartite,hauke2016measuring,lambert2019estimates,lambert2020revealing,yin2019quantum,zheng2015probing,li2013spin,ma2009fisher}, and recently in several experiments~\cite{scheie2021witnessing,laurell2021quantifying,yu2020experimental,gianfrate2020measurement}. The QIG is interesting as a witness of genuine multipartite entanglement~\cite{hyllus2012fisher} that can be probed directly in scattering experiments~\cite{hauke2016measuring}. The QIG also detects the utility of quantum states for quantum metrology~\cite{giovannetti2006quantum,giovannetti2011advances}, allowing for measurement precision beyond the classical bounds of precision~\cite{ou1997fundamental}. Recently, there has been renewed interest in geometrical properties of the quantum state space for their own sake, with Ref.~\cite{erdmenger2020information} noting that non-interacting field theories appear to exhibit a flat state space geometry and Ref.~\cite{zanardi2007bures} emphasizing the relationship between state space curvature and different scaling regimes. A central concept in QIG is the quantum Fisher information (QFI)~\cite{wootters1981statistical,braunstein1994statistical} from which a metric can be derived, the quantum Fisher information metric (QFIM)~\cite{provost1980riemannian}. 

In this work, we investigate the QFIM in the context of the ground states of the spin-1 antiferromagnetic Heisenberg model, with uniaxial anisotropy $D$, and applied field $B_z$
\begin{equation}
  H_0 = \sum_{r} \boldsymbol{S}_r\cdot\boldsymbol{S}_{r+1} + D(S^z_r)^2 + B_z S_r^z,
  \label{eq:DefnSpin1Model}
\end{equation}
a paradigmatic model of low dimensional magnetism for which the ground state at the isotropic point exhibits symmetry protected topology order protected by $\pi$ rotations about any two given spin axes~\cite{chen2011complete}. The ground state is thus gapped, and the resulting ordering can be best characterized as a valence bond solid (VBS)~\cite{affleck2004rigorous} (for a more thorough discussion, see Sec.~\ref{subsec:Spin1Chain}). We show that the ground state entanglement of the VBS phase is enhanced by the presence of small uniaxial anisotropy and is undiminished by the application of homogeneous magnetic fields up to energies equal to the gap. Once the applied field reaches a strength approximately equal to the gap we demonstrate the onset of massive genuine multipartite entanglement as the ground state enters a magnon condensate phase~\cite{sorensen1993large}. Finally, we introduce a method of estimating the state space curvature in the vicinity of the ground state, finding that non entangled regions of the phase diagram appear flat, while entangled regions exhibit positive curvature, and critical regions negative curvature. 

To paraphrase Carlton M. Caves, Hilbert space is a big place~\cite{caves1996quantum}. Consequently, any analysis of QIG begins with specifying some interesting submanifold of states. Throughout this paper, we will be concerned with manifolds that can be generated from an initial state, $\ket{\psi(\xi,0)}$, depending possibly on some other parameters $\xi$, via unitary transformations generated by Hermitian operators of the form,
\begin{equation}
    \hat{\Lambda}_{\boldsymbol{v}^{(\mu)}}  = \sum_r (-1)^r \boldsymbol{v}^{(\mu)}\cdot\hat{\boldsymbol{S}}_r,
    \label{eq:DefnStaggeredMagOp}
\end{equation}
where $\boldsymbol{v}^{(\mu)}\in\mathbb{R}^3$ are vectors that determine the generators $\hat{\Lambda}_{\mu}$, and hence the unitary evolution of the state $\ket{\psi}$ through the Schr\"{o}dinger equation,
\begin{equation}
    -i\partial_{\eta_{\mu}}\ket{\psi} = -i \hat{\Lambda}_{\boldsymbol{v}^{(\mu)}} \ket{\psi}.
\end{equation}
Explicitly, the state $\ket{\psi(\boldsymbol{\eta})}$ is evolved into the state $\ket{\psi(\boldsymbol{\eta}+\dd \eta_{\mu})}$ via,
\begin{equation}
    \ket{\psi(\boldsymbol{\eta} + \dd \eta_{\mu})} = e^{-i\dd \eta_{\mu} \Lambda_{\mu}}\ket{\psi(\boldsymbol{\eta})}
\end{equation}

The distance element between $\ket{\psi(\boldsymbol{\eta})}$ and $\ket{\psi(\boldsymbol{\eta} +\dd \eta_{\mu})}$ is determined by the quantum Fisher information metric (QFIM), $\mathcal{F}_{\mu\nu}$ which, for pure states reduces to the real part of the covariance matrix for the hermitian generators~\cite{provost1980riemannian},
\begin{equation}
    \mathcal{F}_{\mu\nu} \{\psi\} =  \frac{1}{S^2} \Re{\text{Cov}_\psi (\Lambda_\mu,\Lambda_\nu)},
    \label{eq:DefnQFIMatrix}
\end{equation}
(we hereafter assume that $S=1$ and omit the prefactor) with,
\begin{equation}
    \text{Cov}_{\psi}(\Lambda_\mu,\Lambda_\nu) = 
    \qexp{\Lambda_\mu\Lambda_\nu} - \qexp{\Lambda_\mu}\qexp{\Lambda_\nu}.
\end{equation}
For a more detailed discussion of the QFIM see Sec.~\ref{subsec:QuantumFisherInformation}.

The QFIM can used to detect the presence of genuine multipartite entanglement in the state $\ket{\psi}$ via the bound~\cite{hyllus2012fisher,scheie2021witnessing},
\begin{equation}
    \bar{f}\coloneqq\frac{1}{3N}\Tr{\mathcal{F}} > 2m.
    \label{eq:DefnFirstBound}
\end{equation}
for states that are $(m+1)$-partite entangled. A major advantage of the QFIM is that, at finite temperature, it is directly related to the dynamical response~\cite{hauke2016measuring}. Recently, the QFIM has been measured in magnetic systems via dynamical neutron scattering~\cite{scheie2021witnessing,laurell2021quantifying}. The fact that finite temperature entanglement is, at low temperatures, strongly controlled by the entanglement in the ground state~\cite{gabbrielli2018multipartite,lambert2019estimates,menon2023multipartite}, makes investigations of the ground state QFIM relevant to experiments. 

Traditionally, entanglement in many body systems has been explored from the perspective of measures such as the entanglement entropy and R\'{e}nyi entropy~\cite{calabrese2009entanglement} which generally require knowledge of the ground state to evaluate. Such measures have been extensively explored, yielding interesting results in spin chains~\cite{castro2012entanglement,castro2013entanglement}. In particular, much attention has been paid to the area law scaling of the entanglement entropy as a means of characterizing entanglement in many body systems~\cite{eisert2008area}. Despite the enormous theoretical success of this approach, the experimental roadblocks to direct measurement of the entanglement entropy impose a sharp limitation on its utility in this context.  

Entries of the QFI matrix have now been studied in a wide range of models~\cite{yin2019quantum,lambert2020revealing,yang2008fidelity,zheng2015probing,li2013spin,ma2013euler,liu2013quantum,pezze2017multipartite}. These studies have emphasized either parameterizations that are linear in the spin degrees of freedom, or in the case of ~\cite{pezze2017multipartite}, parameterizations using non-local operators that exhibit super extensive scaling in topologically non trivial phases. 
The full QFI matrix that would correspond to these parameterizations has been relatively less explored. 

Using the mean QFI density for a staggered magnetic field, we first construct a map of the multipartite entanglement for the spin-1 antiferromagnetic Heisenberg chain with uniaxial anisotropy, $D$, and applied magnetic field oriented along the $z$ axis, $B_z$ (see Sec.~\ref{subsec:Spin1Chain}).
Subsequently, using the quantum metric, we calculate the volume of a two-dimensional slice of state space parameterized by the orientation of a perturbatively small staggered magnetic field. By taking the ratio of the volume of this slice of state space in the vicinity of the ground state to the volume of a sphere in a flat state space (see ~\ref{sec:Methods} for our definition of flat), we compute the local curvature in the quantum state space. The Haldane phase appears to be characterized by the presence of a large positive curvature, while the trivial insulator phase and N\'{e}el phase appear to be flat.  

In the remainder of the introduction, we expand our discussion of the QFI and its quantitative relationship to GME. We also review the relevant details of the spin-1 antiferromagnetic Heisenberg chain, which we use as a test case. In Sec.~\ref{sec:Methods} we introduce our generalization of the construction in ~\cite{palumbo2018revealing} and the notions of quantum volume and quantum curvature. The results of our analysis of the Heisenberg chain are presented in~\ref{sec:Results}
with concluding remarks given in~\ref{sec:Conclusion}.

\subsection{Spin-1 Heisenberg Chain}\label{subsec:Spin1Chain}

We  focus on the spin-1 antiferromagnetic Heisenberg chain with uniaxial anisotropy, $D$ and applied magnetic field along the $z$ axis, $B_z$,
\begin{equation}
  H_0 = \sum_{r} \boldsymbol{S}_r\cdot\boldsymbol{S}_{r+1} + D(S^z_r)^2 + B_z S_r^z,
\end{equation}
hereafter referred to as the Heisenberg chain. 
It is a well studied model in low dimensional magnetism which has
been extensively studied
~\cite{botet1983ground,botet1983finite,peters2009spin,haldane1983continuum,
sorensen1993large,white2008spectral,so1994equal,lambert2019estimates,
chen2003ground,langari2013ground, pollmann2010entanglement,nakano2022haldane}. 
 The completely isotropic point ($D=B_z=0$) is an example of a phase with symmetry protected topological (SPT) order~\cite{chen2011complete} and exhibits a characteristic doubling of the spectrum of the entanglement Hamiltonian~\cite{pollmann2010entanglement}. We call this phase the isotropic phase or the Haldane phase interchangeably.
The ground state in this phase is a singlet with a bulk gap to a degenerate triplet mode~\cite{haldane1983continuum} for periodic boundary conditions.
In the presence of 
an applied magnetic field, the degeneracy of this mode is lifted, with one (or two depending on the field orientation) 
of the triplet modes diminishing in energy until a lower critical field, $B_z^\text{lower}$ where hybridization
with the ground state singlet induces a phase with long range AFM order that can be interpreted as a BEC phase~\cite{sorensen1993large,zvyagin2007magnetic,giamarchi2008bose}. We use the term BEC or magnon BEC to refer to this phase from this point on. Once the upper critical field $B_z^\text{upper}$ is attained, the per-site magnetization saturates and the spins form a classical paramagnet.
For
large and positive $D$, there is a Gaussian transition to a so called ``Large-D Phase'', which we refer to as the insulator phase, 
and for negative $D$ a transition to a quasi-ordered N\'{e}el
phase~\cite{chen2003ground,langari2013ground}. 

Notice that at the isotropic point, this model commutes with the total magnetization operator along any arbitrary orientation, and that with non-zero anisotropy and applied field, the total magnetization along the $z$ axis is conserved. This will help to motivate our choice of operator in Eq.~\ref{eq:DefnStaggeredMagOp} in the following section.

\subsection{Quantum Fisher Information}\label{subsec:QuantumFisherInformation}

The degree to which two probability distributions may be distinguished from one another in some fixed set of measurements induces a natural notion of distance on the state space~\cite{fisher1922mathematical}. For a classical probability distribution $p(x|\boldsymbol{\eta})$ giving probability of outcome $x$ depending on some set of parameters $\boldsymbol{\eta}$, this notion of distance is quantified by the Fisher-Rao metric~\cite{bengtsson2017geometry}
\begin{equation}
    F_{\mu\nu} \{p(x|\boldsymbol{\eta})\} = \int \frac{1}{p} \pdv{p}{\eta_\mu}\pdv{p}{\eta_\nu} \dd x.
    \label{eq:DefnClassicalFI}
\end{equation}
The quantum generalization holds for both pure~\cite{wootters1981statistical} and mixed~\cite{braunstein1994statistical} states, and is termed quantum Fisher information (QFI) with the density matrix $\hat{\rho}$ taking the place of $p(\omega|\boldsymbol{\eta})$, and the quantum expectation value replacing the integral.
The associated geometrical structure is termed \emph{quantum information geometry} or simply 
\emph{quantum geometry} (see Ref.~\cite{bengtsson2017geometry} for a complete introduction to the subject or Ref.~\cite{lambert2023classical} for a briefer pedagogical introduction). 

The relationship between the QFI and genuine multipartite entanglement is derived in ~\cite{hyllus2012fisher}. We recall again the operator introduced in Eq.~\ref{eq:DefnStaggeredMagOp}, 
\begin{equation}
    \Lambda_{\boldsymbol{v}^{(\mu)}}= \sum_r (-1)^r\boldsymbol{v}^{(\mu)}\cdot\hat{\boldsymbol{S}}_r.
    \label{eq:DefnMagGenerator}
\end{equation}
In the most general case, there may be any site dependence one could imagine for the summand. We are motivated in our choice of the staggered magnetization by the particular nature of the spin-1 Hamiltonian in Eq.~\ref{eq:DefnSpin1Model}. We call Eq.~\ref{eq:DefnMagGenerator} the \emph{generator} of the QFI. In choosing these generators we might, in the most general case, allow the vector $\boldsymbol{v}^{(\mu)}$ to have a site dependence. The relative orientation of the summands has a significant impact on the QFI. In particular, if a generator commutes with the Hamiltonian, then the ground state will be an eigenstate of the generator, and the covariance, and hence the QFI, will be zero. We discuss the implications of this fact for the spin-1 Heisenberg chain more in Sec.~\ref{sec:Methods}. The generator that will detect the greatest QFI is the one for which $[H,\Lambda]$ is maximal. Often at the critical point this will be the most relevant operator\cite{hauke2016measuring} in the renormalization group sense.

At this point it is clear why we choose to measure variants of the staggered magnetization operator defined in Eq.~\ref{eq:DefnStaggeredMagOp}. The spin-1 Heisenberg chain commutes with the uniform magnetization operator, and so the QFI generated by the total magnetization is zero. In terms of the QFI, the fully anti-ferromagnetic generator detects the greatest amount of entanglement ~\cite{lambert2019estimates}. This is easy to see from the equivalence between the QFI and the equal time structure factor, which, for the AFM Heisenberg chain, exhibits a peak at $k=\pi$~\cite{white2008spectral,so1994equal}. 

A natural choice for the vectors $\boldsymbol{v}^{(\mu)}$ is given by $\boldsymbol{v}_1 = \boldsymbol{x}$, $\boldsymbol{v}_2=\boldsymbol{y}$ and $\boldsymbol{v}_3 = \boldsymbol{z}$. From this point on we will simply write $\hat{\Lambda}_{\mu}$ to denote the operator associated with the vector $\boldsymbol{v}^{(\mu)}$ or collectively $\hat{\boldsymbol{\Lambda}} = (\hat{\Lambda}_{1},\hat{\Lambda}_2,\hat{\Lambda}_3)$.
The diagonal entries of the $3\times 3$ QFI matrix associated with the generators corresponding to this choice is given by,
\begin{equation}
    \mathcal{F}_{\mu\mu} = \qexp{\hat{\Lambda}_\mu^2} - \qexp{\hat{\Lambda}_\mu}^2,
    \label{eq:QfiMatrixDiagonal}
\end{equation}
where $\mu \in \{1, 2, 3\}$ correspond to the three orthogonal unit vectors in $\mathbb{R}^3$.

Now, we can consider the QFIM at the state $\ket{\psi}$ associated with the three-dimensional parameterization generated by the vector of operators $\boldsymbol{\Lambda}$. The corresponding QFI matrix, $Q$ (see Eq.~\ref{eq:DefnQFIMatrix}) can be used to define an intensive, \emph{mean QFI density}~\cite{hyllus2012fisher}
\begin{equation}
    \bar{f} = \frac{1}{3N} \Tr(\mathcal{F}),
    \label{eq:DefnQfiDensityTraceQ}
\end{equation}
where $\mathcal{F}$ is the QFI matrix and $N$ is the number of sites. The threshold for genuine multipartite entanglement is then given (in the thermodynamic limit) by $\bar{f} > 2m$. The bound in Eq.~\ref{eq:DefnFirstBound} can be arrived at by considering the operator defined in Eq.~\ref{eq:DefnMagGenerator} and then integrating over all possible orientations of the unit vectors with fixed relative orientations on each site (see Ref.~\cite{hyllus2012fisher} for details).

There is a different way of using the operator $\hat{\Lambda}_{\mu}$ to define a parameterization in state space. Instead of acting directly on the state $\ket{\psi}$ with the unitary operator $\exp{-i\dd \eta_{\mu}\hat{\Lambda}_{\mu}}$, we take $\ket{\psi}$ to be the ground state of a Hamiltonian $\hat{H}_0$, and then perturb this Hamiltonian
\begin{equation}
    \hat{H} = \hat{H}_0 + \eta_{\mu} \hat{\Lambda}_{\mu}
    \label{eq:PerturbedHamiltonian}
\end{equation}
In this context, the distance between the state $\ket{\psi(0)}$ and the state $\ket{\psi( \eta_{\mu})}$ is called the fidelity susceptibility~\cite{zanardi2006ground,yang2008fidelity}. Provided we are away from a level crossing, the state $\ket{\psi(0)}$ is related to the state $\ket{\psi(\eta_{\mu})}$ via the unitary operator,
\begin{equation}
    \hat{U} = \ket{\psi(\eta_{\mu})} \bra{\psi(0)}
\end{equation}
which can be evaluated in perturbation theory provided that $\eta_{\mu}$ is small. The QFIM can still be written in the covariance form of Eq.~\ref{eq:DefnQFIMatrix}, only now instead of the covariance of $\hat{\Lambda}_{\mu}$ we consider the covariance of $\hat{\tilde{\Lambda}}_{\mu}$ defined by,
\begin{equation}
    \hat{\tilde{\Lambda}}_{\mu} = i (\partial_{\eta_{\mu}} \hat{U})\hat{U}^\dagger.
\end{equation}
The resulting metric is denoted by $\tilde{\mathcal{F}}$, and it can be shown that, for a system with energy gap $\epsilon$~\cite{zanardi2006ground}, 
\begin{equation}
    \tilde{\mathcal{F}}_{\mu\nu} \leq \frac{1}{\epsilon} \mathcal{F}_{\mu\nu}.
\end{equation}
For a more thorough discussion of the relationship between $\tilde{\mathcal{F}}_{\mu\nu}$ and $\mathcal{F}_{\mu\nu}$ one can refer to Ref.~\cite{lambert2023classical}. The important point is that the manifold generated by applying a unitary operator and manifold generated by perturbing a parent Hamiltonian are intimately related, with distances between states on the latter bounded by distances between states on the former.

\begin{figure}[tb]
  \centering
  \includegraphics[width=\columnwidth]{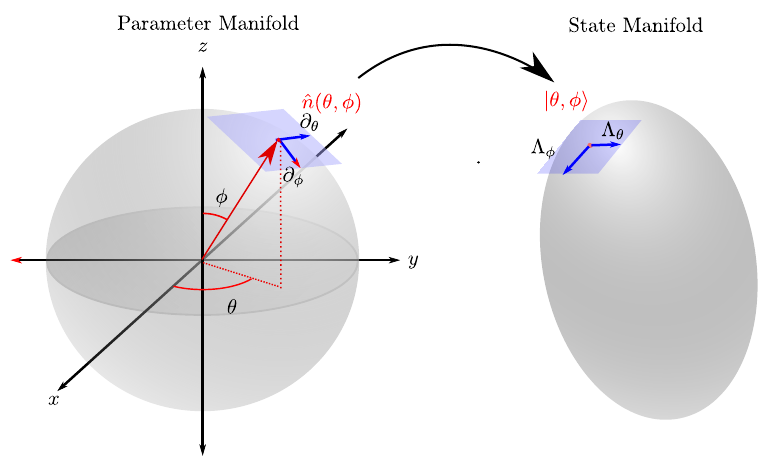}
  \onecolumngrid
  \caption{A schematic of the manifold generated by the applied field
    parameterized by $\{\theta,\phi\}$, with the corresponding ground state given
  by $\ket{\theta,\phi}$. The tangent vector $\partial_\mu$
  becomes the generator $\Lambda_\mu$.}
  \label{fig:SchematicOfManifolds}
\end{figure}

\section{Methods}\label{sec:Methods}
\subsection{Quantum volume and curvature}

\subsubsection{Defining a manifold}
In order to investigate the state space geometry in the vicinity of many body
wavefunctions we consider the two-dimensional sub manifold of the state space that generalizes the manifolds of the form considered in~\cite{yu2020experimental,palumbo2018revealing} to a many-body context. In that work, the manifold we introduce here generated by the operator $\Lambda_{n_r}$ defined in Eq.~\ref{eq:DefnMagGenerator} was examined for just a single site.

We return now to the scenario in Eq.~\ref{eq:PerturbedHamiltonian}, and take $\boldsymbol{\eta} = (h, \theta,\phi)$, 
\begin{equation}
  \hat{H}(\boldsymbol{\xi};h,\theta,\phi) = H_0(\boldsymbol{\xi}) + h\Lambda_{\mathbf{n}}(\theta,\phi),
  \label{eq:DefnHamiltonian}
\end{equation}
where $\boldsymbol{\xi}$ correspond to parameters of the Hamiltonian such as, for example, $D$ and $B_z$ in Eq.~\ref{eq:DefnSpin1Model}.
Infinitesimal shifts in the values of $\boldsymbol{\eta}$ can be generated by the following operators (where, by abuse of notation, we use subscript $\{h,\theta,\phi\}$ as a label),
\begin{subequations}
\begin{align}
    \hat{\Lambda}_{\mathbf{n}} &= \sum_{r} (-1)^r \mathbf{n}\cdot\hat{\mathbf{S}}_{r} \\
    \hat{\Lambda}_{\theta} &= \sum_{r} (-1)^r \mathbf{d}_{\theta} \cdot\hat{\mathbf{S}}_{r} \\
    \hat{\Lambda}_{\phi} &= \sum_{r} (-1)^r \mathbf{d}_{\phi} \cdot\hat{\mathbf{S}}_r
\end{align}
\end{subequations}
where,
\begin{subequations}
    \begin{align}
        \mathbf{d}_{\theta} &= \partial_{\theta}\mathbf{n} \\
        \mathbf{d}_{\phi} &= \partial_{\phi}\mathbf{n}.
    \end{align}
\end{subequations}
For each orientation of $\mathbf{n}$ and value of $h$, there is a corresponding ground state whose geometry can be explored via the metric in state space can be determined using Eq.~\ref{eq:DefnQFIMatrix} (see Fig.~\ref{fig:SchematicOfManifolds}). In the limit of $h\rightarrow 0$ we recover the case described above Eq.~\ref{eq:QfiMatrixDiagonal}.
In state space this manifold can be considered in terms of slices at fixed values of $h$,
\begin{equation}
    \Omega_h=\{\ket{h,\theta,\phi}| (\theta,\phi)\in [0,2\pi)\times [0,\pi]\},  
    \label{eq:DefnSubmanifolds}
\end{equation}
with the state manifold (actually a submanifold) given by $\mathcal{P}=\bigcup_{h>0} \Omega_h$.

Choosing a manifold in state space parameterized by the orientation and strength of a magnetic field makes this study distinct from many other investigations of the many-body quantum state space geometry, which tend to employ state space manifolds parameterized by the driving parameters of the phase transitions (see, for example ~\cite{zanardi2006ground,zanardi2007bures,zanardi2007information,janyszek1989riemannian,venuti2007quantum,kolodrubetz2013classifying,gutierrez2021quantum}, or by examining the geometry of the momentum bands in Fermion models (see~\cite{kemp2021nested} for a very general treatment of $N$ level systems). 

From the metric, the \emph{quantum volume}  of the state space manifold $\Omega_h$ can be calculated,
\begin{equation}
  V(\lambda;h) = \int_{\mathcal{S}} \sqrt{\det \mathcal{F}}\, \dd\theta\dd\phi
  \label{eq:DefnQuantumVolume}
\end{equation}
where $\mathcal{S}=[0,2\pi)\times[0,\pi)$. 
Here we use the word volume to refer to the size of the 2D state space manifold.

In the limit $h\rightarrow 0$ the QFI matrix used to define the mean QFI density in Eq.~\ref{eq:DefnQfiDensityTraceQ} can be recovered by considering perturbations in the $x$, $y$, and $z$ directions as defined by the generators $\Lambda_\mu$ with $\mu\in\{x,y,z\}$ discussed in Sec.~\ref{subsec:QuantumFisherInformation}. The mean QFI density is then computed by evaluating the three connected covariances in Eq.~\ref{eq:QfiMatrixDiagonal} in the ground state of the source Hamiltonian. 

By contrast, when we wish to compute the quantum volume of the 2D submanifold depicted schematically in Fig.~\ref{fig:SchematicOfManifolds}, we evaluate the covariances of the tangent operators $\Lambda_\theta$ and $\Lambda_\phi$ for states that are held at some small, non-zero field $h$ with orientation $(\theta,\phi)$.

\subsubsection{Volumes in Flat Space}

 As an example, consider the quantum metric for a space of $S=1/2$ particles for which there is no energy landscape (i.e. $H_0=0$ in Eq.~\ref{eq:DefnHamiltonian}). We call a space where $H_0=0$ an \emph{empty} space, as there is nothing to distinguish between different spin states except the orientation of the generator. In this case the generator cannot be treated as a perturbation because there is nothing to perturb and the ground state manifold is isomorphic to a sphere. Explicitly for the case of $S=\frac{1}{2}$ the QFI matrix is, 
 \begin{equation}
   \mathcal{F} = 
  Nh^2
  \begin{pmatrix}
      1         &  0 \\ 
     0          &  \sin^2(\phi)
  \end{pmatrix},
\end{equation}
 The resulting quantum volume is,
\begin{equation}
    V_{S=\frac{1}{2}} = 4\pi N h^2.
\end{equation}
(a detailed derivation is given in App.~\ref{sec:AppendixQuantumVolSpinHalf})
this expression can be generalized to the case of spin-$S$
\begin{equation}
  V_\text{P} = 4\pi \left(\frac{ N}{2S}\right) h^2.
  \label{eq:DefnSpinSQuantumVolume}
\end{equation}
(see App.~\ref{sec:AppendixQuantumVol}).
The volume of the quantum paramagnetic has no dependence on the relative orientations of $\hat{n}_r$. Moreover, the volume scales exactly as $h^2$ and thus the local geometry of the empty space is \emph{flat}, with zero curvature. In the classical case, the covariance of the generators of tangential transformations will be zero, since there are no quantum fluctuations. This can be seen explicitly from Eq.~\ref{eq:DefnSpinSQuantumVolume} which goes to zero in the classical limit, $S\rightarrow\infty$. A non-zero quantum volume can be taken to indicate the presence of quantum fluctuations, since this non-zero quantum volume is coming from the fluctuations in the spin expectation value in the direction transverse to the orientation of the generator. Hence, the quantum volume represents the same information as the Heisenberg uncertainty relations for the spin components and is an intrinsic consequence of the relationships between the angular momentum components. We will return to this point in discussing the results for the $S=1$ Heisenberg chain.

\subsubsection{Quantum Curvature}
The notion of state space curvature can be developed by considering the ratio of the quantum volume of the manifold $\Omega_h$ centered on a ground state $\psi$ relative to the quantum volume in an empty space. The \emph{quantum volume ratio},
\begin{align}
  v_h(\psi(h)) &\coloneqq \frac{V(\psi(h))}{V_\text{P}} \nonumber \\
  &= v_0 - b R h^2 + \mathcal{O}(h^3) 
  \label{eq:DefnVolRatio}
\end{align}
may be expanded about small values of $h$, where in the quadratic term determines the curvature of the manifold $\mathcal{P}$.
This is essentially the definition of curvature familiar from classical differential geometry, with the scalar curvature $R$ controlling the degree to which the scaling of volumes is enhanced or suppressed in spaces with negative or positive curvatures, respectively. 
We call $R$ the \emph{quantum curvature}. It is independent of the metric normalization $A$, and should therefore hold for any choice of QFI. The constant $b$ is a positive number that depends only on the dimension of the manifolds used to compute the volumes. For the case of a two-dimensional sphere $b=\frac{1}{12}$. The negative sign can be understood by imagining drawing a circle on a saddle or on a sphere. On the saddle (a surface with negative curvature), the circumference of the circle will be greater than in the Euclidean case, and so the scaling of the circumference with the radius of the circle will be greater than in the case of flat space, with the opposite argument going through the sphere (a surface with positive curvature)~\cite{do1992riemannian}.

When we speak of the quantum volume ratio of a state $\psi$, we mean the quantum volume of the spheroidal shell centered at $\psi$ in the state space.

\subsection{Numerical Methods}

All data was collected using an iDMRG algorithm implemented using the ITensor
library~\cite{itensor}. Simulations were performed with bond dimensions of up to
$2800$ in the vicinity of the critical points, and optimization was run until
truncation errors not exceeding $10^{-9}$ were achieved. In the BEC phase, we found that convergence was challenging, likely owing to the long range nature of the correlations that phase~\cite{zvyagin2007magnetic}. 

Within the iDMRG, correlation functions were measured to a distance of $1000$ sites starting from the center of an effectively infinite chain. These correlations were then used to compute the per-site connected covariances required for the evaluation of the quantum volume.

The sum over $N$ sites is easiest to compute in the iDMRG because we can take advantage of the translation invariance. The covariances involve a double summation over the correlation matrix $C_{r_1,r_2}$,  $\sum_{r_1=1}^N\sum_{r_2=1}^N C_{r_1,r_2}$. In systems with translation invariance we can assume that $C_{r_1,r_2} = C_{|r_1-r_2|}$ and reduce the double sum to $N\sum_{r=1}^N C_{r}$. If we had instead considered a finite system with a boundary, the entire double sum would've been needed. While this is still tractable for non-trivial system sizes when considering the mean QFI density, evaluating the quantum volumes becomes challenging in all but the most symmetrical cases.

In order to compute the integral in Eq.~\ref{eq:DefnQuantumVolume}, we can take advantage of the symmetries of the source Hamiltonian. We derive (see App.~\ref{sec:AppendixSymmetricFormulae}) the following formulae for the case of full rotational symmetry and axial symmetry, 
\begin{widetext}
\begin{subequations}
\begin{align}
   V^{\text{Spherical}}(\lambda;h) &= 4\pi Ah^2\sqrt{\Xi_{xxyy}} \\
    V^\text{Axial}(\lambda;h) &= 2\pi Ah^2 \int_0^\pi 
    \sqrt{\Xi_{xxyy}\cos^2(\phi)\sin^2(\phi)
    -2\Xi_{yyzx}\cos(\phi)\sin^3(\phi) +\Xi_{yyzz}\sin^4(\phi)}\,\,\dd\phi
\end{align}\label{eqs:SymmetricalFormulae}
\end{subequations}
\end{widetext}

\begin{figure}[t]
  \centering
  \includegraphics[width=\columnwidth]{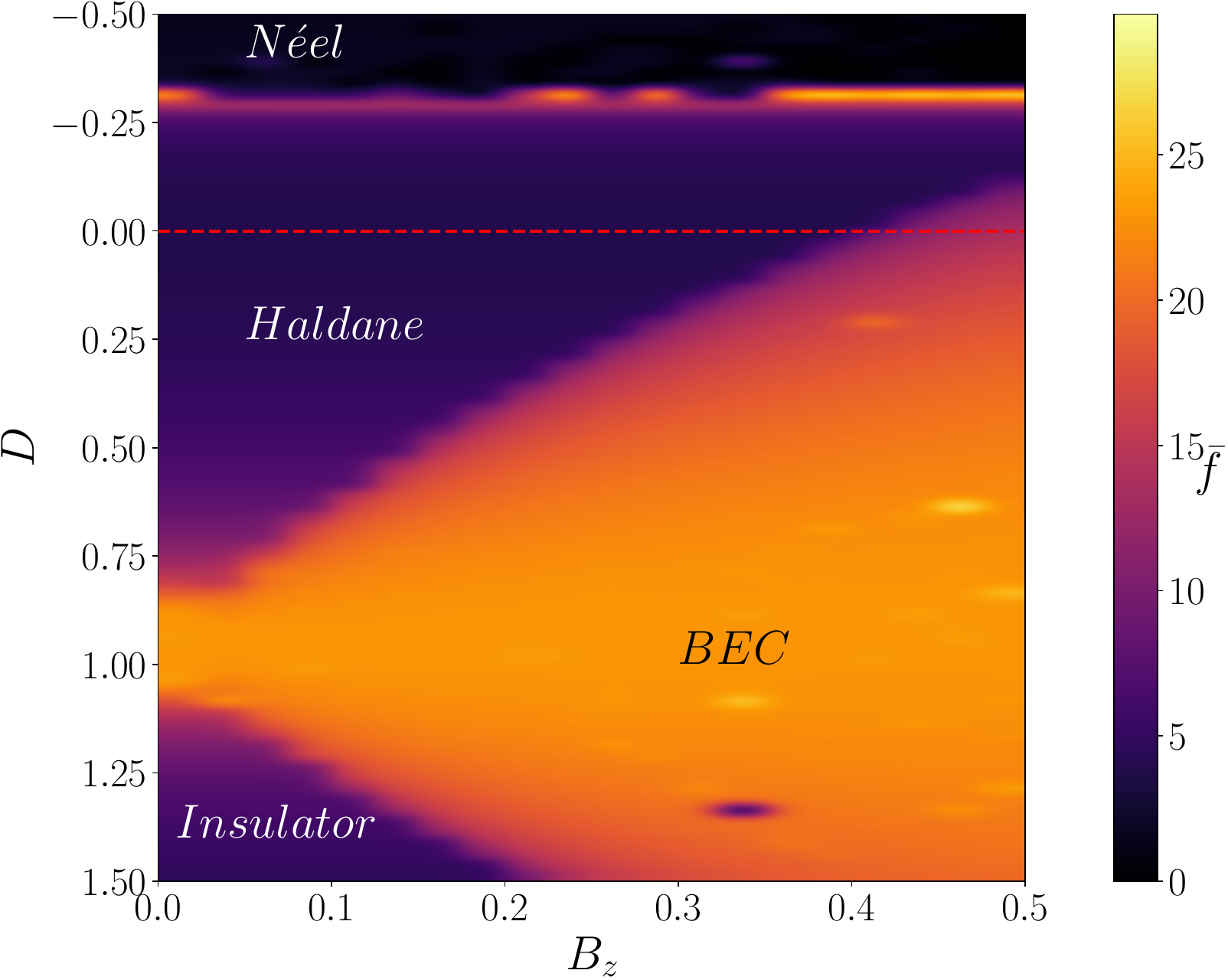}
  \caption{Phase Diagram of the mean QFI generated by the staggered
    magnetization operator. The N\'{e}el and Insulator phases are un-entangled
  away from the critical point. The Magnon BEC phase appears as the critical fan
of the Gaussian critical point separating the Haldane phase from the Insulator
phase. Regions of low entanglement in the BEC phase are numerical artifacts from
failed convergence of the iDMRG. When points do converge, it is with a truncation
error of no more than $10^{-9}$. The red horizontal line indicates the slice that is shown and extended in Fig.~\ref{fig:zeroD_BField_zeromode}}
  \label{fig:D_HZ_PhaseDiagram}
\end{figure}

where in general,
\begin{equation}
    \Xi_{abcd}(\lambda; h,\theta,\phi) = \mathcal{A}_{ab}\mathcal{A}_{cd} - \mathcal{A}_{ac}\mathcal{A}_{bd}.
\end{equation}
Here $\mathcal{A}_{ab}$ is the real part of the correlation matrix, 
\begin{equation}
    \mathcal{C}^{ab} 
    = \sum_{r_1,r_2}
    (-1)^{r_1+r_2}\text{Cov}_{\psi(\theta,\phi)}
    (S_{r_1}^a,S_{r_2}^b),
\end{equation}
given by $\mathcal{A}=\frac{1}{2}(\mathcal{C}+\mathcal{C}^*)$ and the indices $a,b,c,d\in\{x,y,z\}$ are taken in the lab frame. The correlation matrix
will depend on the strength of the perturbing field $h$, and also on the particular values of the Hamiltonian parameters which we here denote as $\boldsymbol{\xi}$. In Eqs.~\ref{eqs:SymmetricalFormulae}, the angular dependence in $(\theta,\phi)$ has been integrated out completely in the spherical case, while in the axial case $\Xi$ will have some dependence on $\phi$.   
In the axial case the integral over $\phi$ can be performed numerically with relatively few integration points. For the data shown in Figs.~\ref{fig:IsoPointSphereRatio}~\ref{fig:quantumVolumeColourPlot} the $\phi$ component was integrated with $50$ equally spaced points. This is relevant for experiments where measurements would be required at a range of field orientations in order to compute the quantum volume.

\section{Results and Discussion}\label{sec:Results}
\subsection{Mean QFI Density}\label{subsec:meanQfiDensity}

Taking Eq.~\ref{eq:DefnSpin1Model} as our source Hamiltonian, the mean QFI density divided by the metric constant was
computed for a patch of the $(D,B_z)$ parameter space with $D\in[-0.5,1.5]$ and
$B_z\in [0, 0.5]$. 
In this section we imagine a $3\times 3$ QFI matrix where the parameterizations are generated by the fields in the $x$, $y$, and $z$ directions in the lab frame as discussed in Sec.~\ref{subsec:QuantumFisherInformation}. The correlation functions are evaluated for the ground state of the source Hamiltonian, taking $h=0$ in Eq.~\ref{eq:DefnHamiltonian}.
The results are given in Fig.~\ref{fig:D_HZ_PhaseDiagram}.
We see robust multipartite entanglement through the Haldane phase, indicating at
minimum bipartite entanglement. We find our results to be consistent with the
low temperature single mode approximation employed by the authors
in~\cite{lambert2019estimates}.

\begin{figure}[t!]
  \centering
  \includegraphics[width=\columnwidth]{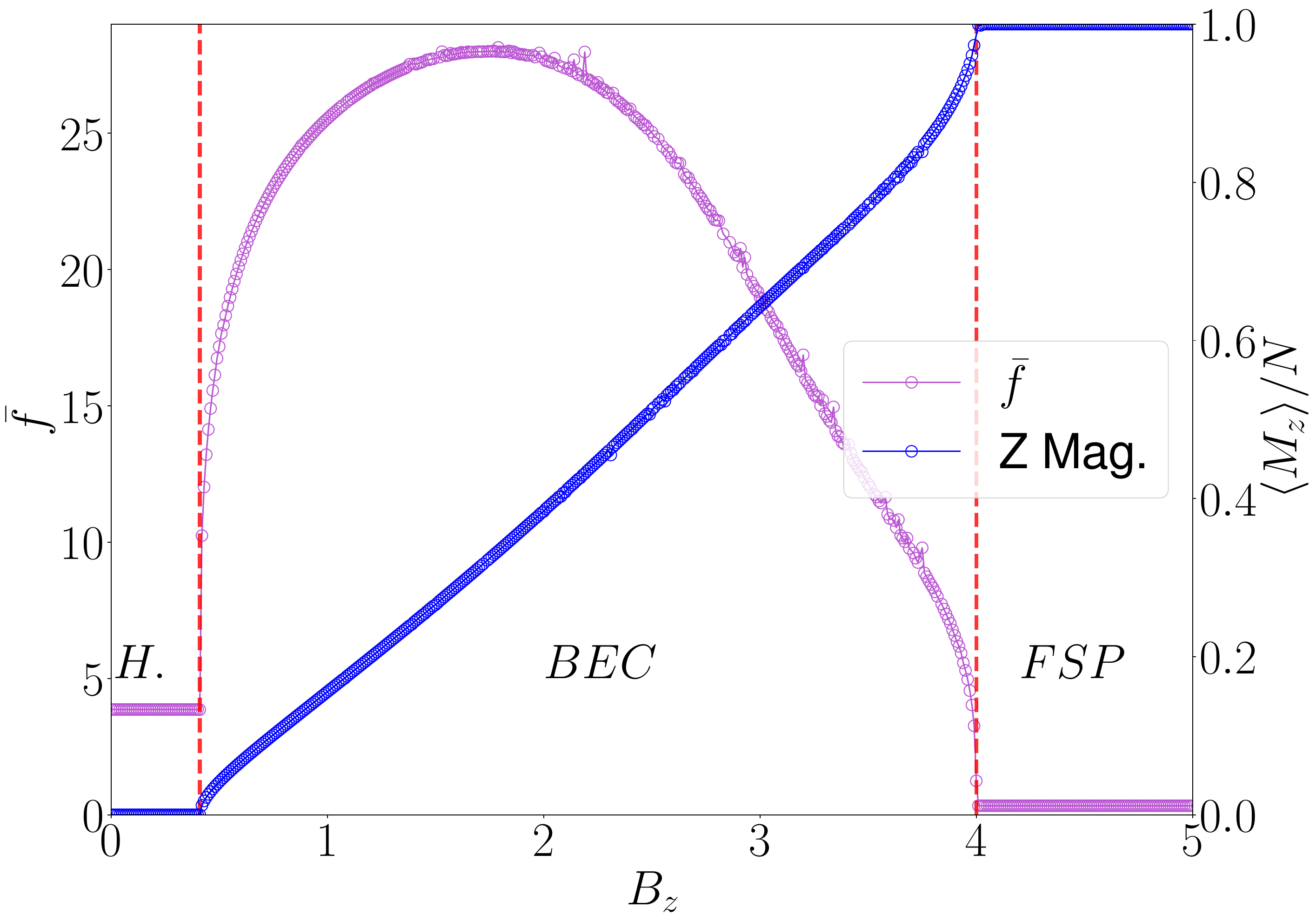}
  \caption{QFI associated with staggered magnetization operator as a function of
  magnetic field in the $z$ direction (magenta curve) and average per site magnetization along the $z$ axis (blue curve). Notice that the mean QFI density is constant in the Haldane phase and nearly zero in the fully spin polarized (FSP) phase.}
  \label{fig:zeroD_BField_zeromode}
\end{figure}

Taking a slice of the phase diagram along $D=0$ (see
Fig.~\ref{fig:zeroD_BField_zeromode}), we see that the multipartite entanglement
of the magnon BEC is substantially higher, and estimate the value of the lower
critical field to be $B_z^\text{lower}\approx 0.41$, which is consistent with
previous studies ~\cite{sorensen1993large} which place the value of $B_z^\text{lower}$ at the Haldane gap ~\cite{nakano2022haldane}. From Fig.~\ref{fig:zeroD_BField_zeromode} the
multipartite entanglement can be seen to peak before falling to the upper critical field
of approximately $B_z^\text{upper}\approx 4$, beyond which the system enters a classical paramagnetic phase for which the mean QFI density is nearly zero and the per site magnetization becomes saturated.

In Fig~\ref{fig:D_HZ_PhaseDiagram} we see that the BEC phase appears as a fan of genuine multipartite entanglement emanating from the Gaussian transition to the insulator phase.
The BEC phase can be reached from either the insulator phase or the Haldane phase. Experimental studies ~\cite{zvyagin2007magnetic} on the material NiCl$_2$-4SC(NH$_2$)$_2$, for which the ground state is in the insulator phase and which exhibits an additional easy axis anisotropy, reveal that the long range correlations in the BEC phase persist up to a finite temperature that grows as a function of the applied field to a maximal point which occurs at approximately the point where the average onsite magnetization assumes a value of $\qexp{M_z}/N\approx0.4$. Owing to the analogous peak in the GME as witnessed by the QFI at the point where the onsite magnetization assumes the values of $\qexp{M_z}/N\approx 0.4$, we speculate that this temperature scale may be related to the massive multipartite entanglement in the ground state. In particular, the temperature curve for the BEC phase in ~\cite{zvyagin2007magnetic} is of a form that is very similar to the curve of genuine multipartite entanglement in Fig.~\ref{fig:zeroD_BField_zeromode}.

Within the Haldane phase the multipartite
entanglement is completely constant as a function of $B_z$, as can be seen in Fig.~\ref{fig:zeroD_BField_zeromode} in the phase labelled H. This indicates that the Haldane phase entanglement is robust to the presence of magnetic fields. Coupled with the already established robustness of the Haldane phase entanglement to finite  temperatures~\cite{lambert2019estimates,gabbrielli2018multipartite}, our observations establish Haldane gap anti-ferromagnets as reliable sources of at least bi-partite entanglement even in the presence of environmental disturbances that tend to decohere entangled states. This has especially important application in light of the application of SPT phases to measurement based quantum computation~\cite{else2012symmetry,raussendorf2022measurement}. 

\begin{figure}[t!]
  \centering
  \includegraphics[width=0.925\columnwidth]{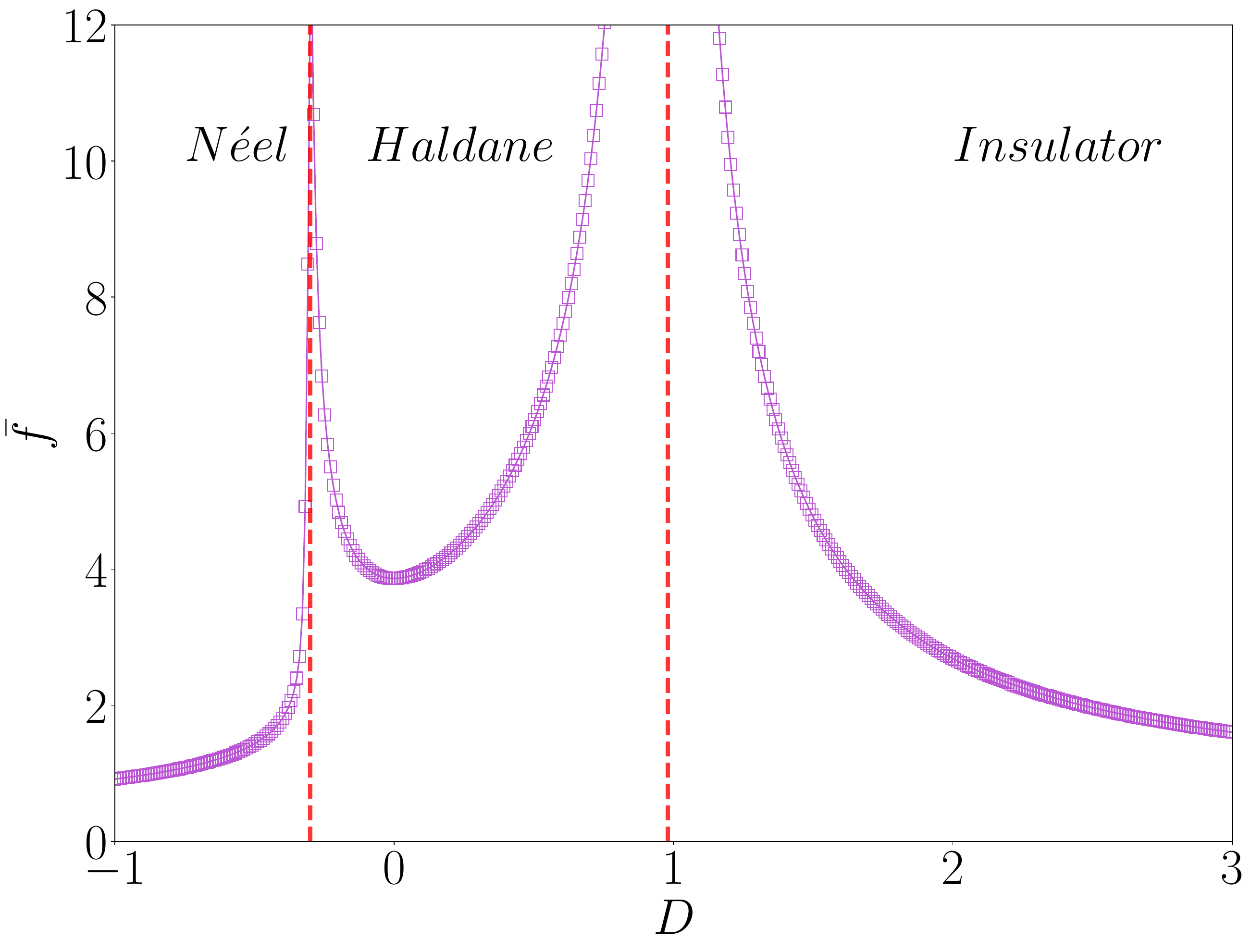}
  \caption{QFI associated with staggered magnetization operator as a function of
  the uniaxial anisotropy. The Haldane phase exhibits the largest amount of multipartite entanglement, which is enhanced by non-zero values of the uniaxial anisotropy.}
  \label{fig:qfiAsFunctionOfD}
\end{figure}

In contrast to a homogeneous applied field, any non-zero uniaxial anisotropy appears to increase the amount of genuine multipartite entanglement, as seen by considering a cut in the state space where $B_z=0$ (see Fig.~\ref{fig:qfiAsFunctionOfD}). Along this cut we can see clear divergences in the GME at both the N\'{e}el and insulator transitions. The Haldane phase exhibits two partite entanglement in the vicinity of the isotropic point where $D=0$, while the other phases are trivial from a quantum perspective, exhibiting zero GME. 

The behaviour of the mean QFI can be contrasted with the fidelity susceptibility, which, for the Gaussian transition in particular, exhibits a less pronounced signal (see Fig. 3 of ~\cite{langari2013ground}). 
While the fidelity susceptibility associated with the Ising transition does exhibit a divergence, this divergence does not establish the entanglement of the critical point, due to the reasons discussed in Sec.~\ref{subsec:QuantumFisherInformation}. 

While the Ising transition was already established as a strongly entangled critical point by the authors in ~\cite{lambert2019estimates}, the observation of entanglement at the Gaussian critical point was not made. This is because the component $Q_{zz}$ of the QFI matrix which was studied by the authors in~\cite{lambert2019estimates} does not diverge at the Gaussian transition, and it is rather the components of the QFI matrix in the directions transverse to the critical point.

\subsection{Quantum Curvature}

In this section we consider the 2D submanifold centered at the ground state of the source Hamiltonian for small, non-zero values of $h$ in Eq.~\ref{eq:DefnHamiltonian}. The correlation functions then in general depend on both $h$ and the orientation of the field $(\theta,\phi)$. Due to the symmetries of the spin-1 chain defined in Eq.~\ref{eq:DefnSpin1Model}, we can integrate out the dependence on $(\theta,\phi)$ at the isotropic point, and the dependence on $\theta$ for the case of non-zero uniaxial anisotropy, using the formulae defined in Eqs.~\ref{eqs:SymmetricalFormulae}. The correlations are then evaluated as functions of $h$ in the case of spherical symmetry, or as functions of $h$ and $\phi$ in the case of axial symmetry. In both cases, the correlations will depend on the value of $D$. The volumes of the 2-spheres for different values of $h$ are sufficient for us to then compute the quantum curvature defined in Eq.~\ref{eq:DefnVolRatio}. If all correlations $\qexp{S_{r_1}^aS_{r_2}^b}$ for a given field strength $h$ and orientation $(\theta,\phi)$ are evaluated with $a,b\in\{x,y,z\}$
in the lab frame then Eq.~\ref{eqs:SymmetricalFormulae} can be evaluated.

\begin{figure}[t]
  \centering
  \includegraphics[width=0.93\columnwidth]{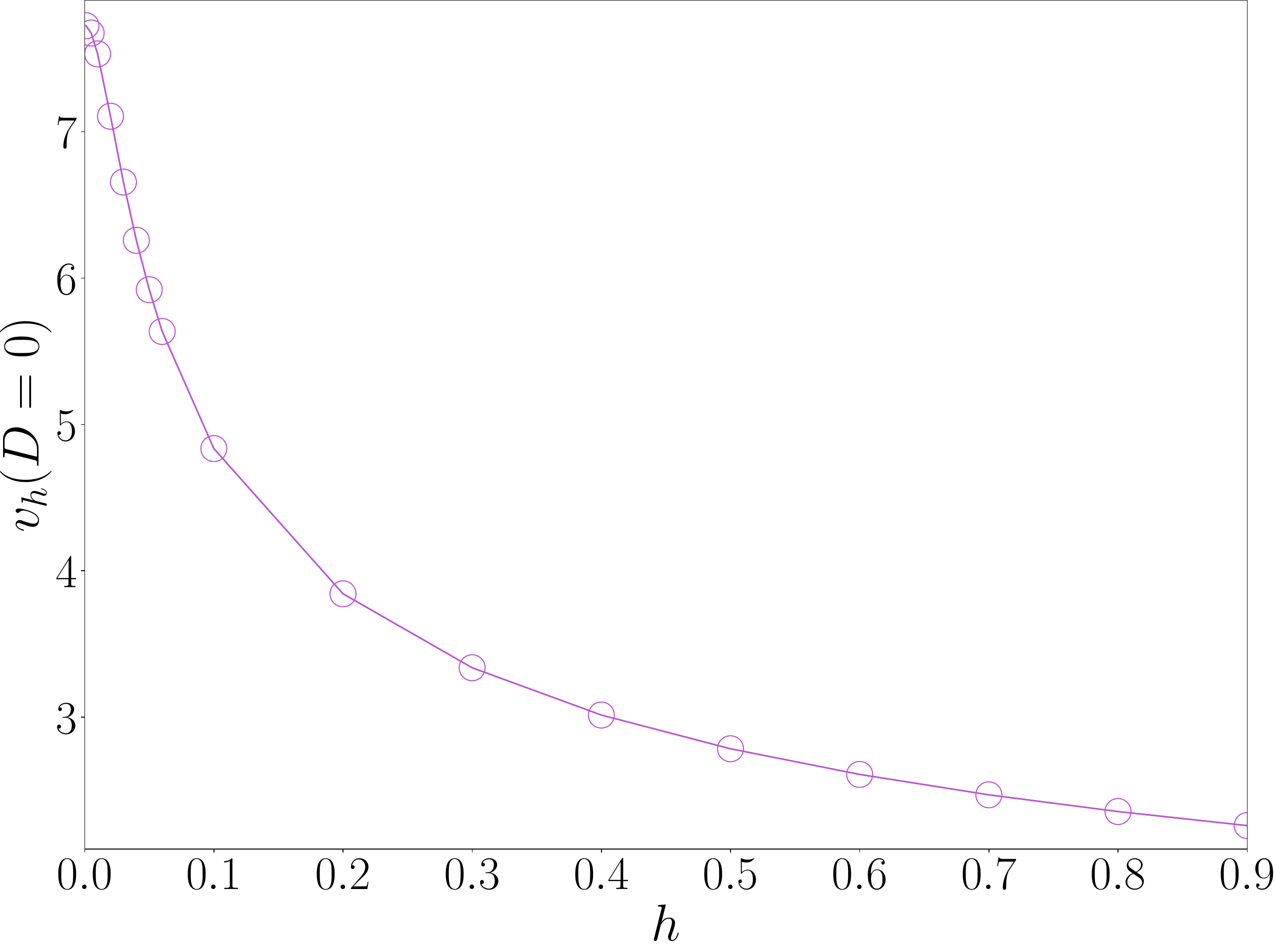}
  \caption{Quantum volume ratio for spheres centered at the isotropic point.}
  \label{fig:IsoPointSphereRatio}
\end{figure}

 We begin at the isotropic point with
$D=B_z=0$ and examine the volume ratios of the ground state of
Eq.~\ref{eq:DefnHamiltonian} as function of $h$ that are taken to be small relative to the exchange coupling. The results of this
calculation are shown in Fig~\ref{fig:IsoPointSphereRatio}. We see that the
volume ratio decreases monotonically, indicating that the manifold $\mathcal{P}$ is asymptotically flat in the limit of large $h$. Near $h=0$ it is clear that
this function is concave down, which, from Eq.~\ref{eq:DefnVolRatio} implies that
the curvature is positive. The volume ratio being larger than 1 indicates the presence of a high degree of quantum fluctuation relative to those implied purely by the uncertainty relation for the spin operators, as discussed in Sec.~\ref{sec:Methods}.

\begin{figure}[t]
  \centering
  \includegraphics[width=\columnwidth]{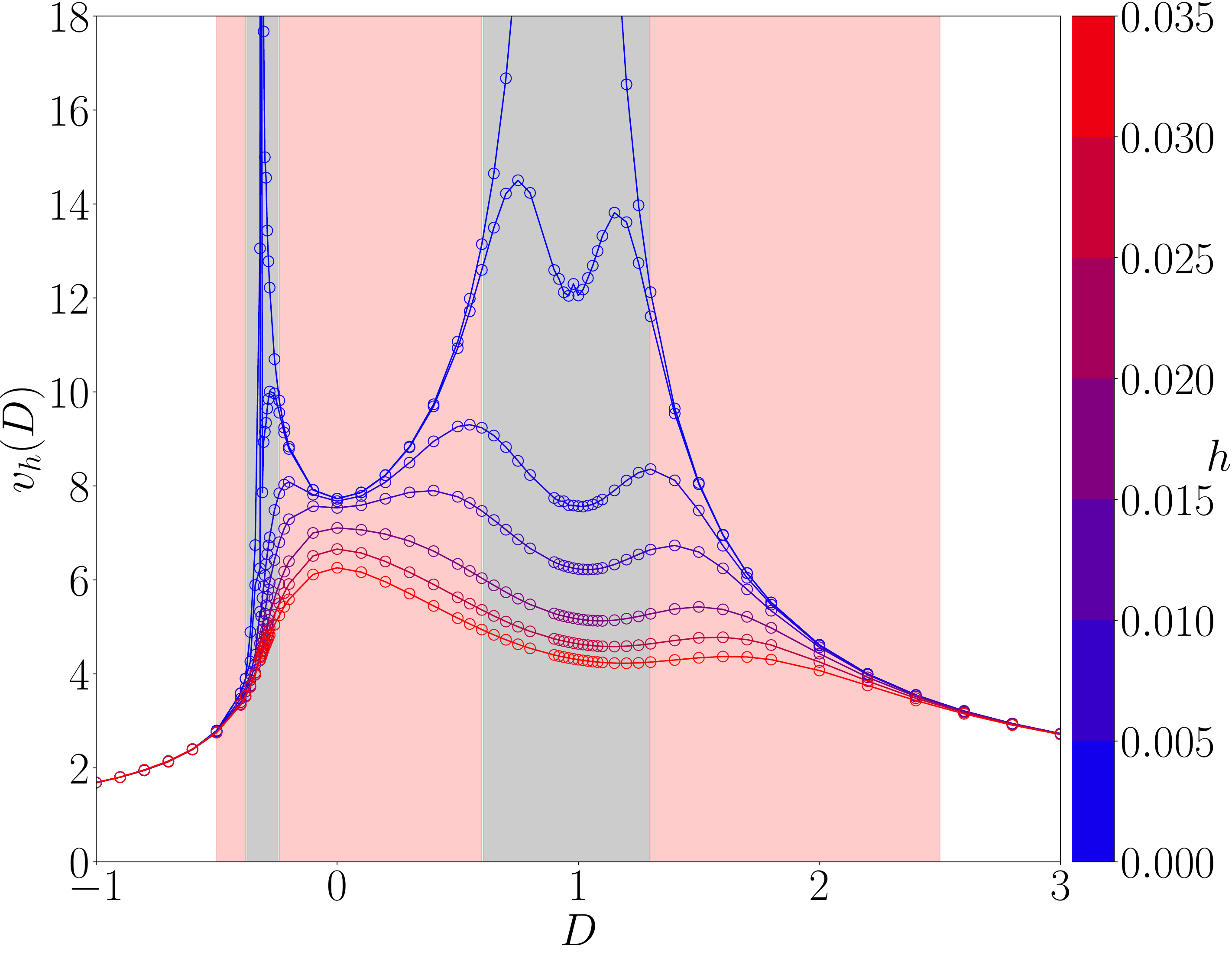}
  \caption{Quantum volume ratio for spheres of perturbatively small radius as a
  function of the uniaxial anisotropy $D$. Sphere sizes (applied fields) range in size from
$h=0.0001$ to $h=0.04$. The red shaded region indicates where the curvature is positive while the grey region indicates where the curvature is negative, approximately corresponding to the onset of the scaling regime.}
  \label{fig:quantumVolumeColourPlot}
\end{figure}

By sweeping across values of $D$ from $-1$ to $3$ we can determine how the curvature of the state space depends on the uniaxial anisotropy. The results of this analysis are shown in
Fig.~\ref{fig:quantumVolumeColourPlot}. Taking the second derivative of the volume ratio as a function of $h$ at $h=0$ gives us the sign of curvature, indicated by the red (positive curvature) and grey (negative curvature) shaded regions in Fig.~\ref{fig:quantumVolumeColourPlot}). We see
that the Haldane phase is characterized by strong positive curvature, while the
N\'{e}el and insulator phases are flat with essentially zero curvature.

Recall from Sec.~\ref{sec:Methods} that the quantum volume quantifies the fluctuations in the spin degrees of freedom. In flat space (i.e. spaces where $H_0=0$), these fluctuations come purely from the Heisenberg uncertainty relations. Hence, the quantum volume in a flat space of $N$ spin-$S$ particles tends to zero in the limit $S\rightarrow\infty$. In the N\'{e}el and Insulator phases, we find a quantum volume ratio $v_h>1$, indicating that these phases exhibit enhanced fluctuations relative to what would be expected from the pure uncertainty relations. 
From Fig.~\ref{fig:qfiAsFunctionOfD}, we see clearly that neither of these phase are entangled, (see the inequality in Eq.~\ref{eq:DefnFirstBound}). Hence, quantum phases might have enhanced fluctuations without necessarily exhibiting any entanglement. 

Despite these enhanced fluctuations, both the N\'{e}el and insulator phases are flat, with quantum volume ratios that are constant as a function of the field $h$. By contrast, the Haldane phase, which exhibits the greatest enhancement in quantum fluctuations of the three, also exhibits a region of large positive curvature, leading into the critical points where we find a sudden, discontinuous jump to a region of negative curvature which we interpret as the onset of the scaling regime. The Haldane phase is the only phase of the three that exhibits any degree of entanglement, and it appears that the quantum curvature tends to grow as the entanglement increases, as seen by the dispersion of the curves at fixed values of $h$ in Fig.~\ref{fig:quantumVolumeColourPlot}. 
The flatness of the insulator phase also establishes that the curvature in the Haldane phase is not simply a consequence of the Haldane gap, as the insulator phase is also gapped but exhibits no quantum curvature.

 It seems clear that the quantum curvature is connected to the presence of entanglement, with the trivial phases appearing completely flat. A phase with no entanglement would then be expected to have zero curvature, except perhaps when the critical point is approached. But it is not clear whether or not the presence of quantum curvature is only incidental in the Haldane phase, and is in fact due to the nearby quantum critical points which exhibit a massive divergence in the amount of GME. 
This would account for the presence of small amounts of positive curvature in the N\'{e}el and insulator phases as the critical points are approached. In this case the positive curvature in the Haldane phase is a consequence of the phases existing in between these two critical points. This introduces the possibility of detecting the presence of quantum critical points even outside of the scaling regime, without the need to directly witness the diverging multipartite entanglement at the critical point. This would be especially useful in inelastic neutron scattering experiments where resolution issues make it effectively impossible to directly measure diverging multipartite entanglement~\cite{scheie2021witnessing}. 
That state space curvature is induced by the phase boundary implies that quantum critical regions have long range effects in the state space, influencing not just the geometry of states in the scaling regime, but also the states that lie in between these critical points. 

It is also possible that state space curvature is generic in the presence of any amount of entanglement, with entanglement and state space curvature sharing a relationship that is somewhat analogous to the relationship between mass and space-time curvature. 

Our calculation of the state space curvature can be contrasted with those of ~\cite{kolodrubetz2013classifying,venuti2007quantum,zanardi2007information} and especially ~\cite{zanardi2007bures,janyszek1989riemannian}. In these studies, it is the curvature corresponding to the ground state manifold parameterized by the driving parameters of the phase transitions themselves that is computed. In particular, ~\cite{zanardi2007bures} finds that the curvature of the manifold parameterized by a uniform field along the $z$ axis and the temperature in the XXZ model reveal different scaling regimes in the vicinity of the quantum critical point. 

This is similar to our observation that the state space curvature becomes negative in the immediate vicinity of the quantum critical point, but our construction pertains to a state space manifold that might be applied to transitions where the magnetic field is not a priori the driving parameter of the phase transition. 
It is also interesting that there is a discontinuous jump from a region of positive state space curvature in the Haldane phase and immediately around the critical points to a region of negative state space curvature in the quantum critical regime, without any intermediate flat region. In the grey regions of Fig.~\ref{fig:quantumVolumeColourPlot}, it is important to emphasize that the form of the quantum metric defined in Eq.~\ref{eq:DefnQFIMatrix} breaks down, as it does not account for the non-adiabatic level crossings that occur at the critical point. A more careful treatment would involve computing the entire spectrum and applying the form of the quantum metric introduced in~\cite{zanardi2006ground}. We leave these questions to future study.

\section{Conclusion}\label{sec:Conclusion}

In this work we have mapped out the phase diagram of the spin-1 antiferromagnetic Heisenberg chain in the space of applied magnetic field and uniaxial anisotropy (both along the $z$ axis). The five phases of the spin-1 Heisenberg chain considered here have all been shown to have substantially different behaviour in terms of the mean QFI density. In particular, we find the both the N\'{e}el and insulator phases exhibit small but non-zero genuine multipartite entanglement relative to the SPT Haldane phase. The entanglement in the SPT phase seems to be greatly enhanced by the presence of unixial anisotropy, and is undiminished under the application of a uniform magnetic field. The fact the multipartite entanglement in the Haldane phase is robust against the application of magnetic fields and exhibits a finite temperature plateau up to energies on the scale of the Haldane gap~\cite{lambert2019estimates,gabbrielli2018multipartite} may have implications from quantum metrology and quantum information, where state exhibiting robust entanglement are used as a resource in various measurement and information processing tasks

Once the applied field reaches a strength on the order of the Haldane gap we find a drastic increase in the multipartite entanglement in the magnon BEC phase. This condensate is found to emanate from the Gaussian quantum critical point that is generated by large positive values of the uniaxial anisotropy, and is similar to the quantum critical fans expected in the finite temperature region above the quantum critical point. In the BEC phase we find a peak multipartite entanglement at $B_z^\text{max}\approx 1.6$. More work should be done to understand the relationship between the massive amount of multipartite entanglement in the ground state and the finite temperature transition points of the BEC phase mapped out in~\cite{zvyagin2007magnetic}. 

By computing the quantum metric corresponding to a state space manifold parameterized by the magnitude and orientation of a small, staggered magnetic field, we introduced the ideas of quantum volume and quantum curvature, and computed both for a range of values of the uniaxial anisotropy. 

Crucially, we have demonstrated that the Haldane phase in this model is characterized by a sizeable, positive quantum curvature, while the N\'{e}el and insulator phases are flat. We have identified two possibilities. First, that the presence of quantum curvature is induced by the presence of quantum critical points, in which case measurement of such curvature may prove applicable as a probe of quantum criticality outside of the scaling regime in neutron scattering experiments. Second, it is possible that massive positive curvature is an intrinsic feature any GME phase. It is important to emphasize that the entanglement in the SPT phase is short range in nature~\cite{chen2011complete}. It is unclear whether or not the state space geometry computed in terms of a linear, local operator would be sensitive to long range entangled phases, where the mean QFI density of local operators does not detect GME~\cite{lambert2020revealing,pezze2017multipartite}.

One final observation is that the mean QFI density along the $B_z=0$ cut shown in Fig.~\ref{fig:qfiAsFunctionOfD}, and the quantum volume ratio in the limit $h\rightarrow 0$, $v_0$ shown as the maximal curve in Fig.~\ref{fig:quantumVolumeColourPlot} appear to be almost exactly proportional to one another. Furthermore, the curvature appears to grow proportionately to the mean QFI density up to the onset of the critical region. This raises the possibility that the GME is a source of curvature in the state space, analogous to the curvature of space time due to the presence of mass-energy. We do not present any rigorous proof of the relationship between $v_0$ and $\bar{f}$ in this work, leaving the problem to future study. 

\acknowledgments
We acknowledge the support of
the Natural Sciences and Engineering Re-
search Council of Canada (NSERC) through Discovery
Grant (No. RGPIN-2017-05759. 
This research was enabled in part by support provided by SHARCNET (sharcnet.ca) and the Digital Research Alliance of Canada (alliancecan.ca).
Part of the numerical
calculations were performed using the ITensor library~\cite{itensor}.
\appendix

\section{Symmetric Formulae for the Quantum Volume}\label{sec:AppendixSymmetricFormulae}

In order to efficiently evaluate Eq.~\ref{eq:DefnQuantumVolume} is crucial that we exploit symmetries of the source Hamiltonian. This would also be the case for experimental explorations of the quantum volume ratio. There are two symmetries that are of particular interest to us. The first is the full rotational symmetry given by the Euler angles $(\theta,\phi)$. 
Recall the rotation operator~\cite{auerbach2012interacting},
\begin{equation}
  R(\theta,\phi) = e^{i\sum_rS^z_r\theta}e^{i\sum_rS^y_r\phi}
\end{equation}
where the gauge angle representing an initial rotation about the $z$ axis is taken to be zero. 
If this operator represents a symmetry of the source Hamiltonian, we may generate the state manifold by applying the rotation to $\mathcal{H}$,
\begin{equation}
  \mathcal{H}(\lambda;h,\theta,\phi) =
  R(\theta,\phi)\mathcal{H}(\lambda;h,0,0)R^{-1}(\theta,\phi)
  \label{eq:DefnTransformLawFullRotationalSymmetry}
\end{equation}
Each element of the quantum metric is given by the real symmetric part of the covariance of the generators,
as defined in Eq.~\ref{eq:DefnQFIMatrix}. Using the particular form of the generator in Eq.~\ref{eq:DefnStaggeredMagOp} and denoting the real symmetric part of the covariance matrix $\mathcal{C}$ gives,
\begin{equation}
    \mathcal{F}_{\mu\nu} = \frac{A}{S^2} h^2d_{\mu}^a d_{\nu}^b \mathcal{C}^{ab}    
\end{equation}
and,
\begin{equation}
    \mathcal{C}^{ab} 
    = \sum_{r_1,r_2}
    (-1)^{r_1+r_2}\text{Cov}_{\psi(\theta,\phi)}
    (S_{r_1}^a,S_{r_2}^b)
\end{equation}
Under a rotation the covariance matrix transforms 
\begin{equation}
  \text{Cov}_{\psi(\theta,\phi)}(S^a_{r_1},S^b_{r_2})
  \rightarrow
  R^{aa'}R^{bb'}\text{Cov}_{\psi(0,0)}(S_{r_1}^{a'},S_{r_2}^{b'})\nonumber
\end{equation}
where Einstein summation is used and, 
\begin{equation}
    R = 
    \begin{pmatrix}
        \cos(\theta)\cos(\phi) & -\sin(\theta) & \cos(\theta)\sin(\phi) \\
        \sin(\theta)\cos(\phi) & \cos(\theta)  & \sin(\theta)\sin(\phi) \\
        \sin(\phi)             & 0             & \cos(\phi)
    \end{pmatrix}.
    \label{eq:DefnRotationMatrix}
\end{equation}
Using this transformation we can extract the angular dependence of the metric for the case of spherical and axial symmetry, and integrate out that dependence completely, leaving
\begin{equation}
   V^{\text{Spherical}}(\boldsymbol{\xi};h) = 4\pi \frac{1}{S^2}h^2\sqrt{\Xi_{xxyy}}
   \label{eq:AppSphericalVol}
\end{equation}
where,
\begin{equation}
    \Xi_{abcd} = \mathcal{A}_{ab}\mathcal{A}_{cd} - \mathcal{A}_{ac}\mathcal{A}_{bd}.
\end{equation}
For the axial case we, without loss of generality, choose the case where the system is symmetric about rotations about the $z$ axis (set $\phi=0$ in Eq.~\ref{eq:DefnRotationMatrix}). In this case we have,
\begin{widetext}
\begin{equation}
    V^\text{Axial}(\boldsymbol{\xi};h) = 2\pi \frac{1}{S^2}h^2 \int_0^\pi 
    \sqrt{\Xi_{xxyy}\cos^2(\phi)\sin^2(\phi)
    -2\Xi_{yyzx}\cos(\phi)\sin^3(\phi) +\Xi_{yyzz}\sin^4(\phi)}\,\,\dd\phi
\end{equation}
\end{widetext}
In the axial case we still have to perform a numerical integration with respect to $\phi$ but good convergence can be achieved using relatively few slices. 

\section{QFI matrix of a flat space for $S=1/2$}
\label{sec:AppendixQuantumVolSpinHalf}

The quantum metric of a flat space can be calculated for N spin-$1/2$ particles by considering the Hamiltonian in Eq.~\ref{eq:DefnHamiltonian} setting $H_0=0$. The tangent 
operators are given by,
\begin{equation}
  \Lambda_\mu = \frac{h}{2}\vec{d}_\mu\cdot\vec{\sigma}
  \label{eq:DefnTangentGenerators}
\end{equation}
Where $d_\mu$ is the derivative of the unit vector along the $\mu$ direction. 

Substituting these definitions into Eq.~\ref{eq:DefnQFIMatrix} (assuming summation over
repeated indices),
\begin{align}
  \mathcal{F}_{\mu\nu} &= Ah^2 d_\mu^a
  d_\nu^b \Re\left(\qexp{\sigma^a\sigma^b} - \qexp{\sigma^a}\qexp{\sigma^b}\right)
  \nonumber \\
  &= A h^2 d_\mu^a d_\nu^b \Re C^{ab}
  \label{eq:DefnQfiMatrixSHalf}
\end{align}
Where we've defined the connected correlation $C^{ab}$. We can use the algebra
of the Pauli matrices to simplify the expression for the correlations somewhat,
\begin{align}
  C^{ab} &= \qexp{\sigma^a\sigma^b} - \qexp{\sigma^a}\qexp{\sigma^b}\nonumber \\
         &= \delta^{ab} + i\epsilon_{abc}\qexp{\sigma^c} -
         \qexp{\sigma^a}\qexp{\sigma^b}
         \label{Eq:CorrelationIdentity}
\end{align}
Now substituting this Eq.~\ref{eq:DefnQfiMatrixSHalf} gives,
\begin{widetext}

\begin{align}
  \mathcal{F}_{\mu\nu} &= 
  h^2
  \Re
  \left(\vec{d}_\mu\cdot\vec{d}_\nu
  +
  i(\vec{d}_\mu\times\vec{d}_\nu)\cdot\qexp{\vec{\sigma}}
  - d_\mu^a d_\nu^b \qexp{\sigma^a}\qexp{\sigma^b}
  \right)
  \label{eq:MetricWithExpectationValues}
\end{align}
\end{widetext}
Now we must evaluate the expectation values of the Pauli matrices. Recall that
these are themselves functions of $(\theta,\phi)$. 
For the two level case we
can compute the ground state explicitly,
\begin{subequations}
  \begin{align}
    \ket{\lambda_-^{\Delta=0}} &=
    \begin{pmatrix}
      \sin(\frac{\phi}{2}) \\
      -\cos(\frac{\phi}{2})e^{i\theta}
    \end{pmatrix}
  \end{align}
\end{subequations}
The expectation value of the Pauli vector in the ground state is then,
\begin{equation}
  \qexp{\vec{\sigma}}
  = 
  \begin{pmatrix}
  -\cos(\theta)\sin(\phi) \\
  -\sin(\theta)\sin(\phi) \\
  -\cos(\phi)
  \end{pmatrix}
  =
  -\hat{n}(\theta,\phi)
\end{equation}
The metric in Eq.~\ref{eq:MetricWithExpectationValues} is now given by,
\begin{equation}
  \mathcal{F}_{\mu\nu}
  =h^2
  \Re\left(
  \vec{d}_{\mu}\cdot\vec{d}_\nu - i(\vec{d}_\mu\times\vec{d}_\nu)\cdot\hat{n}
  \right)
\end{equation}
In this form, evaluating the components of the metric proves to be a
straightforward exercise
\begin{equation}
  \mathcal{F} = 
  h^2
  \begin{pmatrix}
      1         &  0 \\ 
      0 &  \sin^2(\phi)
  \end{pmatrix}
\end{equation}
Generalizing this expression to $N$ spins amounts to multiplying by $N$
\begin{equation}
  \mathcal{F} = 
  Nh^2
  \begin{pmatrix}
      1         &  0 \\ 
      0         &  \sin^2(\phi)
  \end{pmatrix}
\end{equation}
since the connected correlations between sites is zero by construction. Notice that this expression does not depend on the relative orientation of the neighbouring sites.
For $S=\frac{1}{2}$, the quantum volume (surface area of a 2D sphere in the spin Hilbert space) is then,
\begin{equation}
    V_{S=\frac{1}{2}} = 4\pi Nh^2
    \label{eq:QuantumVolSpinHalf}
\end{equation}

\section{Quantum volume of a flat
space for spin-S}\label{sec:AppendixQuantumVol}

In computing to quantum volume ratio, we want to generalize the volume computed in Eq.~\ref{eq:QuantumVolSpinHalf} to the case of spin-S. This is easy to do by employing the formula for the spherically symmetric quantum volume given in Eq.~\ref{eq:AppSphericalVol}

Using Eq.~\ref{eq:AppSphericalVol} we can compute the quantum volume of a paramagnet. Taking $H_0=0$ in Eq.~\ref{eq:DefnHamiltonian} and denoted the ground state of the spin-S paramagnet as $\ket{S,-S}$, we can use the raising and lowering operators,
\begin{subequations}
\begin{align}
    S^+\ket{S,m} &= \sqrt{S(S+1)-m(m+1)}\ket{S,m+1} \nonumber \\
    S^-\ket{S,m} &= \sqrt{S(S+1)-m(m-1)}\ket{S,m-1} \nonumber
\end{align}
\end{subequations}
For the paramagnet the connected correlation between  is exactly zero, so we can evaluate only the onsite correlations in the directions transverse to the orientation of the field (which we take to be $z$). The expectation values in $x$ and in $y$ are the same, so we only present the calculation of the former. For the state that is fully polarized down along the $z$ axis, the only contribution to $(S^x)^2$ is given by,
\begin{align}
    \mathcal{C}^{xx} = \frac{1}{4}\bra{S,-S}S^-S^+\ket{S,-S} &= \frac{S}{2}
\end{align}
Substituting this into Eq.~\ref{eq:AppSphericalVol} gives,
\begin{equation}
   V_P = 4\pi \left(\frac{N}{2S}\right)  h^2
\end{equation}
where we have grouped the spin, S, and number of sites $N$ along with the factor of $\frac{1}{2}$ in order to emphasize the familiar form of the volume of a sphere. The factor $N$ is coming from the fact that $\mathcal{C}^{ab}$ in $\Xi^{abcd}$ contains a sum over $N$ sites, which for the case of $H_0=0$ scales linearly in $N$ (since the connected correlations between sites are zero). We can check and see that for $S=\frac{1}{2}$ we recover Eq.~\ref{eq:QuantumVolSpinHalf}.
\bibliography{ref}

\end{document}